# Carbon nanotube neurotransistors with ambipolar memory and learning functions

Ertürk Enver Yildirim[1], Luis-Antonio Panes-Ruiz[1] ORCID ID: 0000-0002-3007-8840, Pratyaksh Yemulwar[1], Ebru Cihan[1] ORCID ID: 0000-0002-1747-3838, Bergoi Ibarlucea[1]* ORCID ID: 0000-0002-9899-1409, Gianaurelio Cuniberti[1]* ORCID ID: 0000-0002-6574-7848

[1]Institute for Materials Science and Max Bergmann Center for Biomaterials, TUD Dresden University of Technology, Dresden (Germany)

*Bergoi Ibarlucea: bergoi.ibarlucea@tu-dresden.de; Gianaurelio Cuniberti: gianaurelio.cuniberti@tu-dresden.de

**Abstract**

In recent years, neuromorphic computing has gained attention as a promising approach to enhance computing efficiency. Among existing approaches, neurotransistors have emerged as a particularly promising option as they accurately represent neuron structure, integrating the plasticity of synapses along with that of the neuronal membrane. An ambipolar character could offer designers more flexibility in customizing the charge flow to construct circuits of higher complexity. We propose a novel design for an ambipolar neuromorphic transistor, utilizing carbon nanotubes as the semiconducting channel and an ion-doped sol-gel as the polarizable gate dielectric. Due to its tunability and high dielectric constant, the sol-gel effectively modulates the conductivity of nanotubes, leading to efficient and controllable short-term potentiation and depression. Experimental results indicate that the proposed design achieves reliable and tunable synaptic responses with low power consumption. Our findings suggest that the method can potentially provide an efficient solution for realizing more adaptable cognitive computing systems.

**Impact:**

The huge amount of data generated by the current society makes it necessary to explore new computing methods with higher efficiency to overcome the bottleneck formed between data storage and processing tasks. Neuromorphic computing aims at emulating the functioning of our brain, which performs both tasks utilizing the same hardware. Here, we propose ambipolar field-effect transistors (FETs) based on carbon nanotubes with a polarizable gate dielectric, capable of providing memory functions reminiscent of neuronal synapses, at both polarities of the device. The ambipolar characteristic doubles the possibilities of previously demonstrated neurotransistors. The short-term and ambipolar behavior of the device can find its place in novel applications in the future. Machine learning-enabled gas sensing is an excellent example, where real-time processing of large amounts of data is beneficial. In addition, interaction with oxidative and reductive gases will result in dual responses due to the ambipolarity of the transistor, along with the possibility of storing the sensing data.

**Introduction**

The demand for high hardware performance has increased in recent years due to the rise of data-intensive computing applications, which require features such as low access latency, large bandwidth, high capacity, low cost, and the ability to perform artificial intelligence (AI) tasks. However, big data poses additional challenges, such as high energy consumption and limited memory bandwidth. Traditional computer architecture has limitations that affect its performance, such as the von Neumann bottleneck. This bottleneck arises from the separation of processor and memory, requiring data to be transferred between them through a single bus, limiting the overall performance of the system. To address these challenges, neuromorphic computing research is being conducted, aiming to develop computing systems that mimic the way the human brain processes information.[1] This field includes research on neuromorphic materials and devices, circuits, algorithms, and applications. Neuromorphic computing is considered as a potential solution to the challenges faced by conventional computers and is projected to grow rapidly in the coming years, with an estimated market size of $22 billion by 2035.[2]

To advance neuromorphic computing and engineering, it is crucial to explore new materials and devices that can disruptively improve the power efficiency and scalability of current solutions. Most of the reported works have focused on the function and plasticity of biological synapses as key elements in the interlacing between neurons.[3–5] However, the neuron as a whole is also a critical element taking part in memory processes through its intrinsic plasticity[6] by non-linearly integrating the received information in the cell membrane and generating new signals for other postsynaptic neurons. In contrast with many of the previously mentioned synaptic devices, transistors as three-terminal devices resemble with better fidelity the structure of neurons,[7] with a signal that travels over long distance from an input to an output (source and drain) and which is modulated on its way by the input of at least another terminal (gate).

The use of transistor-based artificial neurons and synapses has gathered significant interest due to their ability to emulate complex synaptic functions, perform learning operations and signal transmissions synchronously, and utilize multi-terminal geometries to process information in a non-linear manner, switch and amplify electric signals like biological neurons.[8] More specifically, ion-gated transistors utilize the movement of ions in the vicinity of the semiconductor surface to imitate the dynamic behavior of biological synapses and neurons. These transistors can effectively replicate synaptic weight modulation through ion electrochemical doping at the dielectric/channel interface. Furthermore, the multi-terminal design of the ion-gated neuromorphic transistor allows neurons to replicate their information processing capabilities.[9] However, these devices suffer from slow switching speeds due to the inherent slow speed of ion movement. For example, our previously reported work[10] proposed an ion-gated transistor where the memory effect relied on the use of a polarizable ion-doped sol-gel interfaced with the semiconductor channel as a dielectric gate. The polarization was based on ionic drift and diffusion in the porous sol-gel material under modulation of presynaptic gate input. Derivatives of sol-gel materials have been observed to exhibit high dielectric constant, high breaking strength, and low dielectric loss, making them appealing candidates for use in energy storage devices and electronics.[11,12] Adjusting various parameters, including precursor composition, solvent, processing conditions, and post-treatment techniques enables tailoring the overall properties of the sol-gel to meet specific application demands. In the aforementioned work, despite the speed limitation, the device showed a non-linear sigmoidal potentiation of time-dependent inputs, which emulated the dynamics of neuronal membrane integration. The device architecture with multiple gate inputs as synaptic inputs and the transistor output as an axon terminal emulated the information processing of a true neuron. Experimental results demonstrated the non-linear information processing ability based on the intrinsic plasticity of a neurotransistor, which can be controlled to generate and reach faster or slower spike thresholds to control the rate of the spike. Furthermore, the semiconductor channel of the device was based on silicon nanowires. These, together with other 1D, 2D, or organic materials, are in a non-mature state but offer potentially promising approaches. These options may offer extended functionalities and present new opportunities for flexible electronics or 3D integration.[13]

Most transistors present unipolar charge transport, limiting the learning capabilities. Some strategies can be used to achieve ambipolarity, such as using thin films of semiconductor polymer blends, incorporating materials with different (p- or n-type) behavior.[14] However, high material complexity results in longer fabrication times and higher costs due to the additional steps required to deposit and pattern more materials.[15] For this reason, it is preferable to follow a simpler strategy implementing a single component with intrinsic ambipolar characteristics. FETs based on semiconducting single-walled carbon nanotubes (sc-SWCNTs) are appropriate candidates. Their transfer characteristics present a V- or U-shape with a transition from the hole to the electron conduction region around the Dirac point, holding great promise for ambipolar neuromorphic performance. The ambipolar behavior of sc-SWCNTs is primarily due to their narrow bandgap, enabling the injection of both carrier types as the gate voltage is raised above or below the flat band voltage, provided that the work function of the source/drain electrodes is similar. The fabrication of CNT-FETs is also simpler compared to that of silicon nanowire FETs, where etching, doping, annealing and oxide growth steps are required among others.[16] Our group recently demonstrated that an automated dielectrophoretic alignment methodology was possible to achieve multiple highly uniform CNT-based devices in a single chip and with wafer-scale fabrication compatibility.[17] The existence of commercial inks for printing technologies enables industrial fabrication of large-area electronics, even on flexible devices.

In this research work, CNTs are coated with a polarizable sol-gel silicate film containing metal ions, resulting in the fabrication of ambipolar neurotransistors (Figure 1). The use of sol-gel material offers a costs-effective option compatible with current semiconductor processes.[18] Additionally, sol-gel-derived porous silica films have shown low dielectric constant and loss as wel as high mechanical flexibility due to their porous nature.[19] The proposed scheme allows emulation of the inherent plasticity observed in neurons. By polarizing the gate material, a short-term memory effect is induced, leading to ambipolar memory behavior that can occur in both branches of the current-voltage characteristics. This combined approach incorporates the advantages associated with ambipolar neurotransistors, including the unique mechanical, electrical, thermal, and chemical properties of carbon nanotubes.

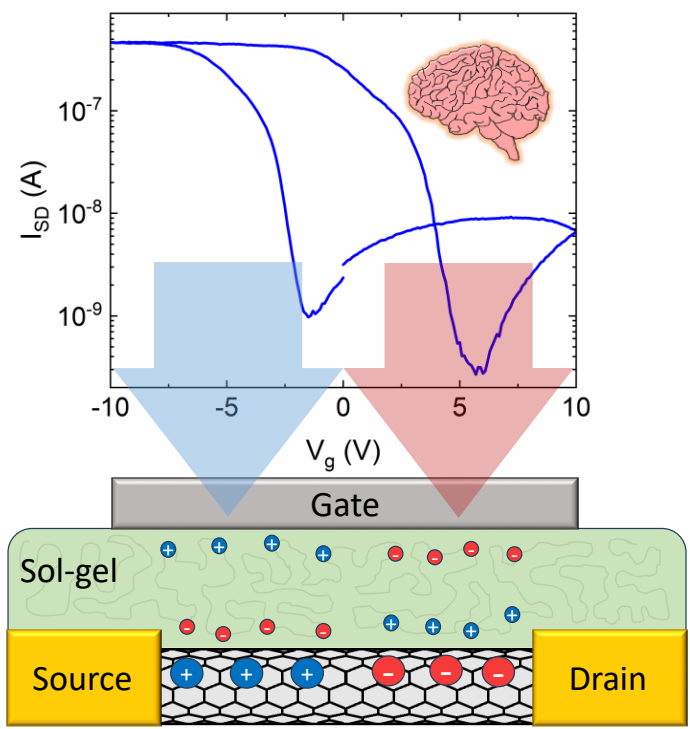

**Figure 1.** Concept of the current work. An ambipolar field-effect transistor based on carbon nanotubes is modified with a sol-gel dielectric gate material loaded with mobile metal ions. Polarization of the sol-gel layer upon gate potential modulation is the base for memory functions, which can happen in the two conduction regions (holes and electrons) of the device.

## Results

*Sol-gel modified CNT-FET characterization*

CNT-based neurotransistors were fabricated by depositing a metal-ion doped sol-gel film on a silicon wafer chip containing thin film gold electrodes bridged by carbon nanotubes, followed by the evaporation of a thin silver film as the gate electrode (see experimental section for details). The mobility of copper and nickel ions in the porous sol-gel in response to the applied gate potential is the main principle of the memory effects of the neurotransistor. The performance of the neurotransistor is influenced by the structure and concentration of metal dopants in the film. Due to the random distribution of metal ions in the sol-gel during gelation, the polarization of the film is zero. However, applying a positive or negative bias across the dielectric film causes the ions to redistribute under the influence of the electric field. Although the ions diffuse back to a random distribution once the bias is removed, the diffusion rate is slower than the previous drift caused by the applied voltage, resulting in a memory effect. When a bias is applied for a prolonged period (around 100 milliseconds), positive and negative ions accumulate and produce a field-effect on the transistor channel. With a sufficiently strong effective field, a dipole moment is induced, leading to a steady increase in current until saturation is reached. In our previous work,[10] it was observed that the sol-gel film acted as a dielectric material whose dielectric constant changed significantly with the concentration of metal dopants at frequencies below 100 Hz. However, the dielectric constant remained constant and was no longer affected by the ionic concentration at frequencies above 100 Hz. The incorporation of the MTMS precursor into the sol-gel film was aimed at enhancing its neuromorphic properties. The addition of organic precursors can also have an impact on the microstructure and properties of the resultant coating, including adhesion, hardness, and optical characteristics. MTMS precursor has been demonstrated to increase the pore density while decreasing the pore volume, thereby increasing the porosity of the gate material.[20,21] However, the introduction of pores compromises the dielectric properties of the material by reducing its dielectric constant. The extent of this effect depends on several factors, such as MTMS concentration, type and concentration of other precursors, reaction conditions (temperature and pH), and post-treatment processes. Therefore, the effect of MTMS on the porosity of sol-gel materials should be evaluated on a case-by-case basis. The MTMS/TMOS ratio was selected based on a prior investigation which demonstrated that the 3:2 proportion exhibited the lowest leaching potential.[22]

The thickness of the sol-gel gate film was measured using scanning electron microscopy (SEM) and was determined to be in the range of 2.6 - 2.8 μm, as depicted in Figure 2a. The surface morphology of the deposited films was examined using atomic force microscopy (AFM), and Figure 2b displays the observed results. The depth of the pores within the films ranged between 17-26 nm, while the width of the pores varied from 115 nm to 175 nm.

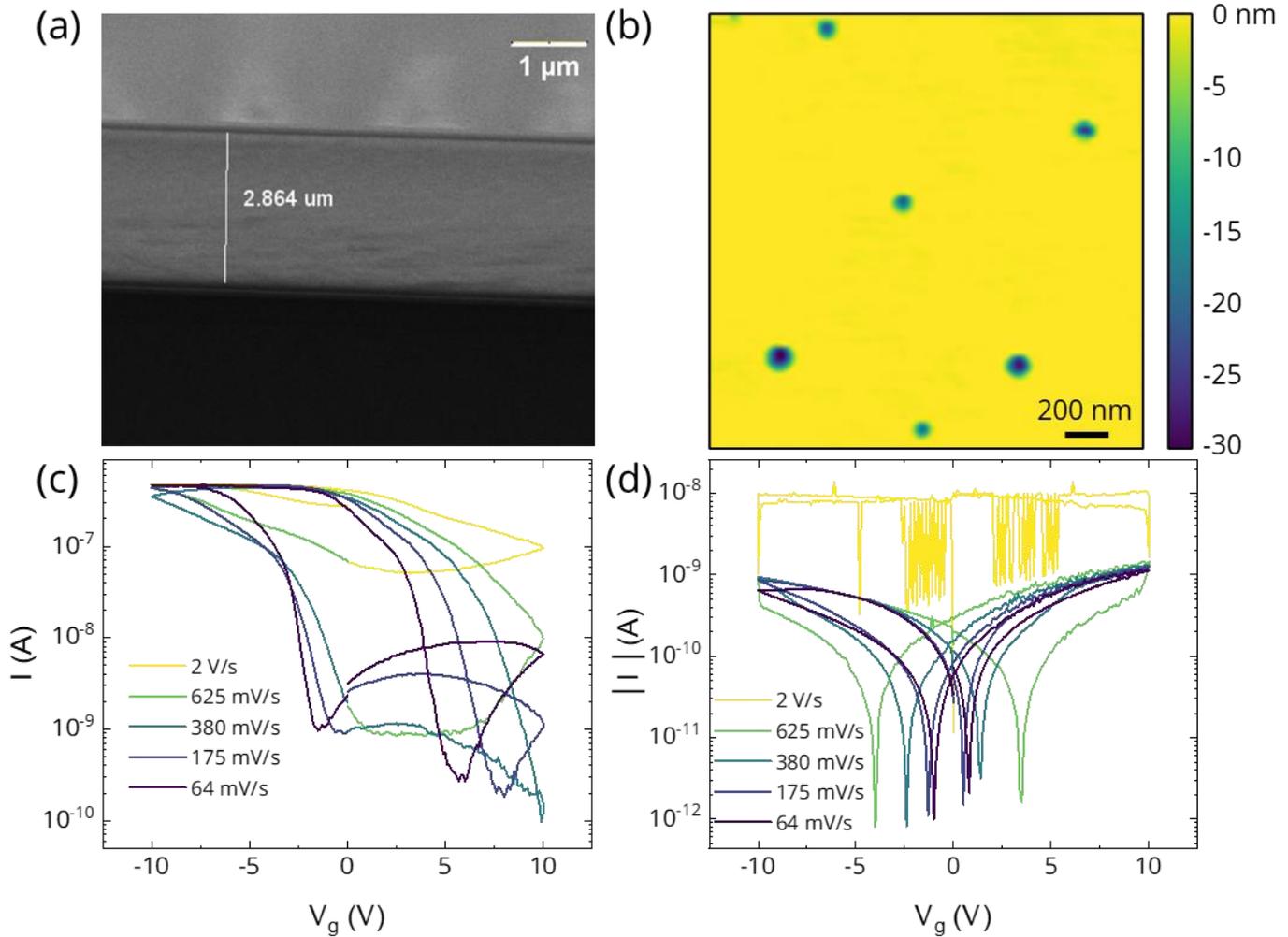

**Figure 2.** Characterization of the sol-gel modified CNT-FET. (a) SEM image showing the thickness of the sol-gel. (b) AFM image of the surface morphology of the sol-gel. (c) Transfer characteristics of the sol-gel modified CNT-FET at different voltage scan speeds. The gate leakage currents at the same speeds are shown in (d).

Figure 2c illustrates the transfer characteristics of the sol-gel modified CNT-FET at different voltage sweeping speeds. At high sweeping rates, the transfer characteristics exhibit a low ON/OFF current ratio of 10 with unipolar behavior. Conversely, when voltage sweeping rates are slow, a high ON/OFF ratio of $10^3$ is observed, accompanied by slight ambipolarity. The ratio of hole conduction to electron conduction follows an order of $10^2$. Leakage current, depicted in Figure 2d, was approximately $10^{-9}$ A and its behavior varied with different voltage sweeping speeds. The leakage current reveals the presence of a polarization switch within the porous gate material, demonstrating the short-term retention of the polarization. This behavior is analogous to that of a lossy dielectric material with free charge carriers or defects.[23] The variations in results obtained with different voltage sweeping speeds can be attributed to slower ion dynamics within sol-gel thin films compared to ferrroic or electronic ordering effects, as well as the presence of reductant Miller capacitances.[24,25] Specifically, the low mobility of ionic movement leads to increased ON/OFF current ratios and manifestation of ambipolar behavior when using lower voltage sweeping speeds. This effect is also observed in leakage current, where the hysteresis of voltage values at which current minima occur depends on the scan rate. This phenomenon is reminiscent of the hysteresis observed in memristive nanodevices, where fast scan rates do not allow sufficient time for the recovery of mobile charges along thin nanoscale semiconductors.[26] Moreover, the sol-gel modified CNT-FET demonstrates comparable gate coupling to ion-sensitive FETs,[27,28] as evidenced by the highest subthreshold slope measured at 120 mV/dec with a voltage sweeping speed of 64 mV/s.

Enhanced gate coupling can significantly improve the ambipolar behavior of CNT-FETs. To achieve better coupling in the future, it is crucial to obtain a thinner sol-gel film. Various methods can be employed for this purpose, including longer plasma treatment durations, different deposition conditions such as adjusting baking temperatures, spin-coating speed and accelerations, as well as varying proportions of solvents in the precursor mixture, among others.

In our experimental efforts, we explored higher water or isopropanol ratios and increased HCl concentrations, which resulted in the production of thinner sol-gels (approximately 234 nm thick, as depicted in Figure S1 of Supplementary Information). However, the corresponding transfer characteristics exhibited a reduced ON/OFF ratio and hysteresis (Figure S2), rendering them unsuitable for memory applications. Additionally, we observed the absence of a distinct electron conduction branch. This could potentially be attributed to the decrease in dielectric properties caused by the increased presence of pores, as illustrated in the AFM image in Figure S3. Furthermore, raising the baking temperature proved to be impractical since temperatures exceeding 100°C caused cracking of the sol-gel (Figure S4).

An alternative approach was explored to enhance gate coupling by modifying the duration of air plasma treatment. Air plasma has the potential to increase the hydrophilicity of CNTs, thereby improving the adhesion of sol-gel to the CNT surface.[29] However, it can also hinder ambipolarity by reducing electron conduction and increasing hole conduction through the generation of oxygen defects on the CNT surface.[30,31] Fourier Transform Infrared Spectroscopy (FTIR) was employed to analyze samples treated with air plasma durations for up to 10 seconds before sol-gel deposition, and the results are illustrated in Figure S5. The observed transmittance in the range of 1680 to 1800 inverse centimeters (1/cm) is attributed to C=O stretching, 1600 to 1680 (1/cm) to C=C stretching, approximately 2150 (1/cm) to C=C=O stretching, around 2349 (1/cm) to O=C=O stretching, and 3500-3780 (1/cm) to O-H stretching. The findings reveal that air plasma treatment rendered the $SiO_2$ surface with CNTs more hydrophilic, evident from the increased OH stretching, thereby promoting improved adhesion of the precursor solution to the surface. However, this process also resulted in the formation of oxygen defects in CNTs, which could potentially impede ambipolarity. These observations are consistent with prior reports.[29] Moreover, longer plasma treatments were found to cause the removal of CNTs from the surface, leading to a decrease in conductance. Notably, the absence of plasma treatment resulted in the absence of gate modulation, likely due to inadequate adhesion of the sol-gel to the substrate. Additionally, overlap between the gate and drain/source electrodes (Figure S6a) could give rise to unnecessary superfluous charging of Miller capacitances.

To mitigate the formation of Miller capacitances, samples with 6 µm SU8-5 passivation were prepared (as depicted in Figure S6b). However, sol-gel deposition on top of the SU8-5 induced defects due to the rough surface of the passivation layer (Figure S7). Furthermore, the viscosity of the sol-gel resulted in thicker gate films along the edges of the SU8-5 layer, leading to decreased gate coupling.

*Short-term potentiation in the p-type region*

To explore the influence of various pulse amplitudes on learning behavior in the p-type region, we employed pulses with a width and time interval of 100 ms each, resulting in a frequency of 5 Hz. The findings presented in Figure 3a demonstrate that higher pulse amplitudes correspond to a greater increase in drain current ($I_{ds}$).

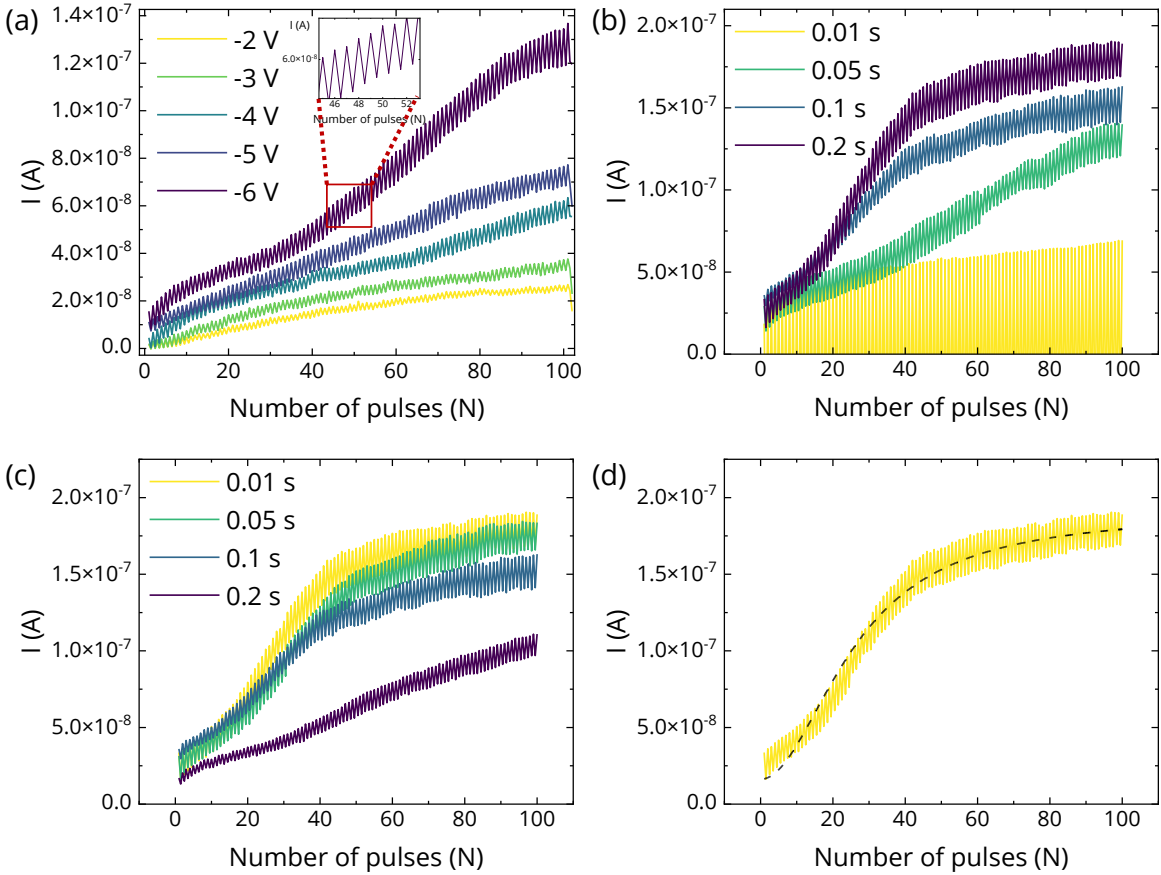

**Figure 3.** Short term plasticity (STP) of the sol-gel modified CNT-FETs in the p-type region. (a) STP with different pulse amplitudes (from -2 V to -6 V) with 100 ms pulse width and 100 ms pulse interval. (b) STP at -6 V and 100 ms time intervals, pulse width varying between 0.01 s and 0.2 s. (c) STP at -6 V and 100 ms pulse width, time intervals varying between 0.01 and 0.2 s. (d) Sigmoid-function-fitted current corresponding to the input pulses with a pulse amplitude of -6 V, a pulse width of 100 ms, and a time interval of 10 ms.

To compare the effect of pulse widths on learning behavior in the p-type region, the pulse amplitude was fixed at -6 V, and the time interval was held constant at 100 ms. Figure 3b shows that longer pulse widths lead to increased potentiation, which is commonly observed in ionic-based neuromorphic devices.[9,10] On the other hand, very short pulse widths yield a greater change in capacitive current, which is assumed to be due to the width being insufficient for capacitance relaxation at the gate film. To compare the effects of different time intervals between subsequent pulses, the pulse width was maintained at a constant value of 100 ms and the pulse amplitude was fixed at -6 V. The measurements presented in Figure 3c revealed that shorter time intervals led to higher degree of potentiation. Notably, the pre-synaptic spike-timing-dependent plasticity (STDP) of the short-term plasticity (STP) behavior of the neuromorphic transistor exhibited greater sensitivity to time intervals compared to pulse widths in millisecond ranges.

Interestingly, the augmentation of $I_{ds}$ exhibited a transition from a linear learning curve to a sigmoidal one, depending on pulse width, amplitude and time interval. This phenomenon is shown in Figure 3d and bears significance as it is frequently used as an activation function of artificial neurons due to its continuous differentiability, making it beneficial for gradient-based optimization algorithms such as backpropagation.[32]

Based on the conducted analysis, it was observed that the fabricated neuromorphic transistors demonstrated enhanced potentiation in the p-type region under specific conditions. Particularly, higher levels of potentiation were noted in correlation with increased applied voltages, broader pulse widths, and reduced intervals between subsequent pulses.

*Short-term potentiation in the n-type region*

The transfer characteristics of neurotransistors have demonstrated that hole conduction surpasses electron conduction. The subthreshold slope in the n-type region is higher, adversely affecting the behavior of STP by lowering learning rates. It was necessary to apply a positive resetting pulse (10 V) due to depolarization to the hole conduction region that occurred when no gate voltage was applied (see experimental section).

To examine the impact of various pulse amplitudes, pulses with a width and interval of 100 ms, corresponding to a frequency of 5 Hz, were employed. The analysis results depicted in Figure 4a unveiled that, akin to the p-type region, higher pulse amplitudes in the n-type region led to a greater increase in $I_{ds}$. However, we observed that the dependence of the learning rate on voltage changes was less pronounced in the n-type region. Moreover, we discovered that higher voltages were necessary to induce significant alterations in learning rate, which is suboptimal due to increased energy consumption. This drawback poses challenges for ambipolar behavior when different voltages are utilized for the p-type and n-type regions, as well as for neuromorphic computing, owing to heightened energy requirements.

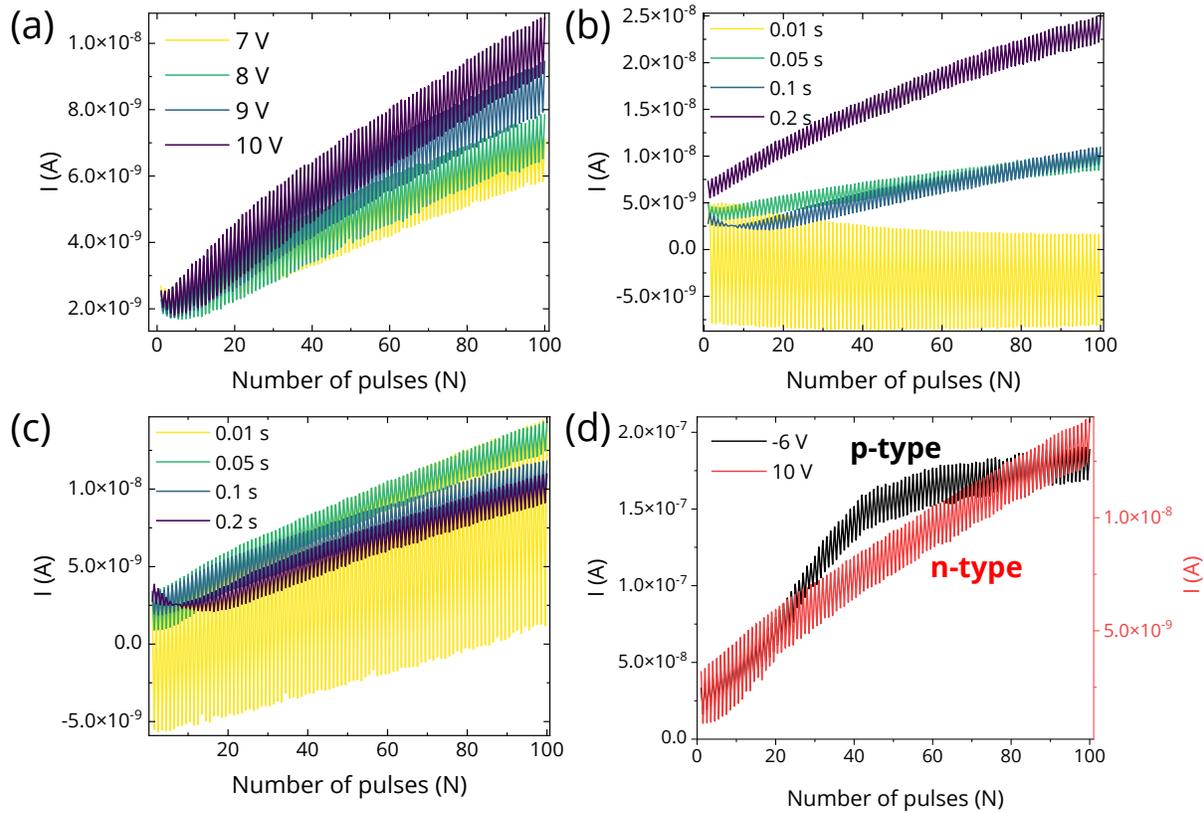

**Figure 4.** Short term plasticity (STP) of the sol-gel modified CNT-FETs in the n-type region. (a) STP with different pulse amplitudes (from 7 V to 10 V) with 100 ms pulse width and 100 ms pulse interval. (b) STP at 10 V and 100 ms time intervals, varying pulse width between 0.01 s and 0.2 s. (c) STP at 10 V and 100 ms pulse width, time intervals varying between 0.01 and 0.2 s. (d) Comparison of learning curve in p-type and n-type modes, showing sigmoidal shape only for the p-type. Conditions of the p-type mode: -6 V, 100 ms pulse width, 10 ms time interval. Conditions of the n-type mode: 10 V, 100 ms pulse width, 50 ms time interval.

To evaluate the impact of different pulse widths, we maintained a fixed time interval of 100 ms and a pulse amplitude of 10 V. Our analysis of Figure 4b revealed that, similar to the p-type region, longer pulse widths resulted in increased potentiation. Additionally, to investigate the influence of varying time intervals between subsequent pulses, we kept a constant pulse width of 100 ms and a pulse amplitude of 10 V. The measurements presented in Figure 4c demonstrated that the time intervals had minimal effect on potentiation in the n-type region. However, we observed that the neurotransistor was unable to sustain 10 ms-time intervals, unlike the p-type region. This discrepancy could be attributed to the faster relaxation rate of negatively held polarization in the sol-gel-derived gate material within the n-type region. Notably, our findings indicate that in the n-type region, the pre-synaptic STDP of STP behavior exhibited greater sensitivity to pulse widths compared to time intervals, contrary to the observations in the p-type region. This can be attributed to the different drift and diffusion times of the positive and negative ions in the sol-gel.

The conducted analysis revealed that neurotransistors exhibit higher potentiation in the n-type region of ambipolar transistors under specific conditions, such as higher applied voltage levels and wider pulse widths. However, learning behavior demonstrated lower sensitivity to time intervals. It is important to note that no discernible transition to sigmoidal learning behavior was observed (Figure 4d). Additionally, the learning rate was lower in all cases when compared to the p-type region.

*Forgetting and erasing in the p-type region*

For the forgetting, we first generated a learning curve, followed by a period of pulse absence. Various negative pulse amplitudes were utilized while keeping the pulse width and the time interval constant at 100 ms. After 100 pulses, only $I_{ds}$ was measured without applying any gate potential. The results (Figure 5a) demonstrate that a higher degree of potentiation correlates with a longer duration for potentiation to be forgotten. The forgetting rate is initially rapid until it reaches a baseline where the current potentiation exhibits a sigmoidal change. Subsequently, it slows down significantly. Consequently, complete forgetting may take several minutes to hours, while forgetting up to a threshold of less than 50% may occur within seconds. However, in none of the cases, does the signal return to the original value. We attribute this to the fact that the CNT-FET is not in OFF state when no gate voltage is applied, and a gate voltage is required to truly set it off. This phenomenon may be due to the remanent polarization within the silicate gate material that is characterized by the random distribution of ions and defects, or the charge trapping caused by CNTs. On the contrary, when positive pulses of 6 V with a pulse width of 100 ms were applied, a much faster short-term depression (STD) was observed, as shown in Figure 5b. Notably, depression can lead to a current level lower than the initial state. Achieving the same relaxed state of the system presents challenges due to the tunable threshold voltage and device ambipolarity, requiring different pulse amplitudes and widths for different polarization levels.

To examine the erasing rates at different erasing voltages, we employed positive gate voltage pulses of various amplitudes with a width and time interval of 100 ms after a learning period. Since it is difficult to achieve exactly the same potentiation level, we compared the erasing voltages by normalizing the $I_{ds}$ values to the maximum potentiation level. The results (Figure 5c) indicate that higher erasing pulse amplitudes correspond to higher erasing rates, similar to the potentiation process. Furthermore, Figure 5d suggests that higher levels of potentiation initially lead to higher erasing rates, followed by slower erase rates after reaching the sigmoidal change baseline.

To investigate the re-learning after forgetting behavior, the samples were first subjected to learning pulses, consisting of 100 consecutive pulses with a -6 V amplitude, 100 ms pulse width, and a fixed 100 ms time interval between subsequent pulses. This was followed by a period of no gate potential to undergo forgetting. Re-learning was then initiated after three different forgetting times: 2 s, 5 s, and 10 s (indicated by yellow, green, and purple arrows in Figure 5e, respectively). The results demonstrate that re-learning after forgetting occurs at a faster rate (slope) compared to the initial potentiation. After the forgotten information is re-learned and reaches a similar current value, the learning rate gradually slows down and returns to normal values.

To assess the effects of erasing on re-learning behavior, we conducted experiments involving a sequence of positive 6 V pulses, corresponding to erasing for a specific duration, followed by -6 V pulses with a 100 ms pulse width and time interval for re-learning. The impact of erasing duration on the re-learning effect was investigated using durations of 2 s, 5 s, and 10 s, as illustrated in Figure 5f. Our results indicate that, due to the rapid erasing rate compared to forgetting, it takes a longer time to recover erased information. Erasing beyond the point where the sigmoidal curve model is exceeded and the threshold point is reached poses a significant challenge for the utilization of neurotransistors. Thus, the optimal positive pulse amplitude, pulse width, and interval time must be carefully selected to achieve the desired erasing behavior.

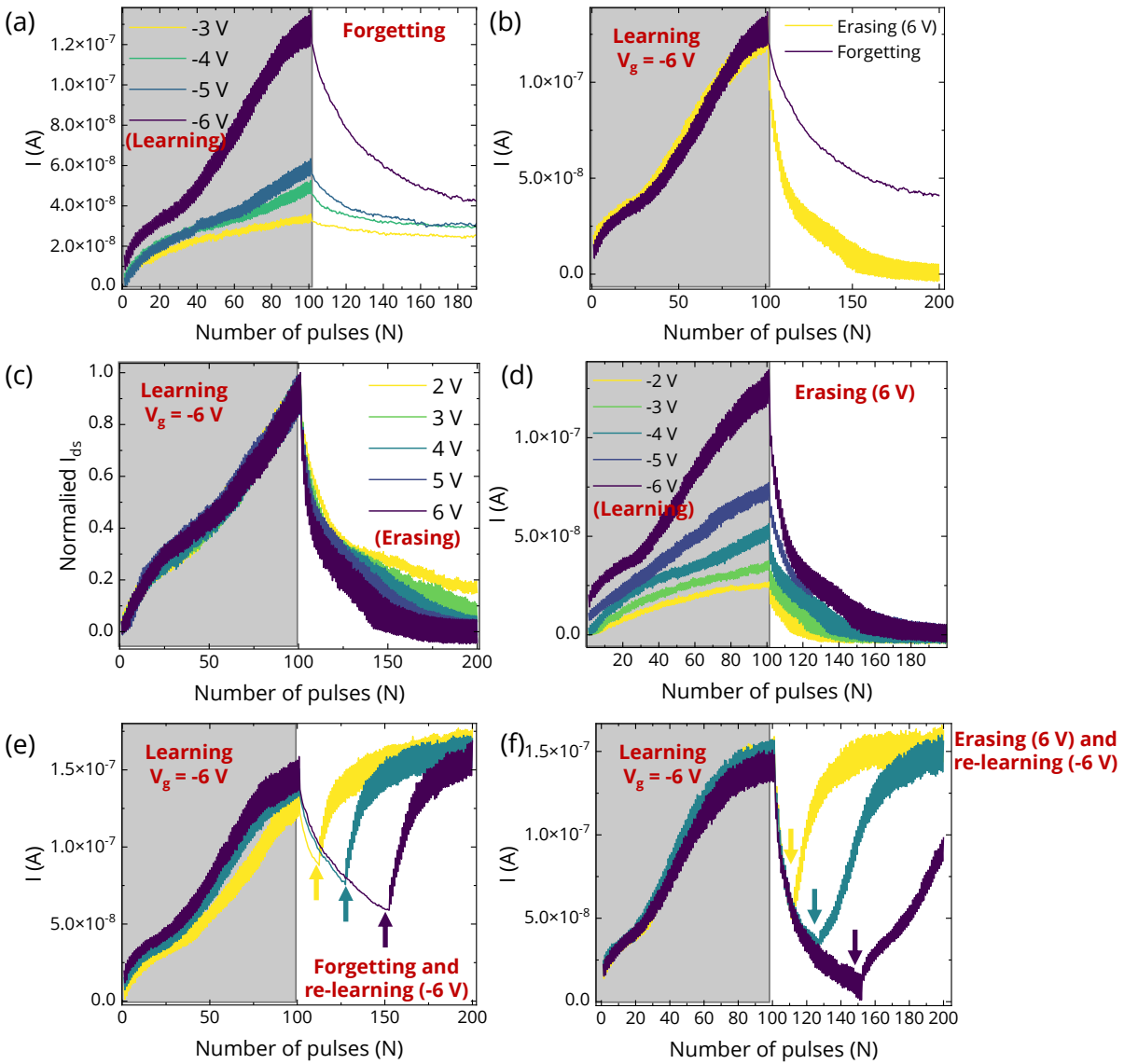

**Figure 5.** Forgetting and erasing in the p-type region of the CNT-FET. (a) Short-term potentiation at different gate voltages followed by a forgetting period. (b) Short-term potentiation at -6 V followed by forgetting and erasing ($V_g = 6$ V). (c) Erasing rates at different erasing gate voltages after a common learning period. (d) Erasing rates at fixed erasing gate voltage (6 V) after different learning processes. (e) Re-learning after three different forgetting times (yellow, green, and purple arrows indicate re-learning after 2s, 5 s, and 10 s of forgetting time, respectively). (f) Re-learning after three different erasing times (yellow, green, and purple arrows indicate re-learning after 2s, 5 s, and 10 s of forgetting time, respectively).

Figure S8 reveals that the erasing rate is amplified by wider pulse widths and shorter time intervals between consecutive pulses, mirroring the behavior observed in STP within the p-type region. Furthermore, similar to STP, a reduction in pulse width results in an increase in the swing of $I_{ds}$ during erasing, which is attributed to capacitive charging effects.

*Forgetting and erasing in the n-type region*

We explored the forgetting behavior in the n-type region of the fabricated neurotransistors by conducting a series of experiments utilizing pulses with varying positive amplitudes. Throughout these experiments, we maintained a consistent pulse width and time interval of 100 ms. After an initial set of 100 pulses, we focused solely on measuring the $I_{ds}$ current. The results (Figure 6a) show a positive correlation between the duration of potentiation and its magnitude. The forgetting rate remained constant in the n-type region regardless of the previously utilized learning potential, in stark contrast to the p-type region where it exhibited variability. Furthermore, our observations indicate that the learned information can be forgotten to the extent of returning to the baseline. Surprisingly, we also discovered that the neurotransistor can transform into a hole conduction region even in the absence of a gate voltage. We noticed that this inherent switch becomes more prominent at the lowest levels of potentiation (using 2 V or 3 V, as shown in Figure 6a). This intriguing phenomenon suggests the presence of an inherent depolarization effect (Figure 6e), reminiscent of ferroelectric memories.[33] It is this depolarization that can ultimately modify the conduction region of the neurotransistor. However, this outcome poses a significant challenge for neuromorphic computing, as it may result in the generation of false information if the processes of potentiation, forgetting, and information refreshment are not carefully optimized. Furthermore, following the polarity switch, we observed an abrupt firing, potentially laying the groundwork for an integrate-and-fire model. These findings underscore the importance of optimizing potentiation and forgetting times.

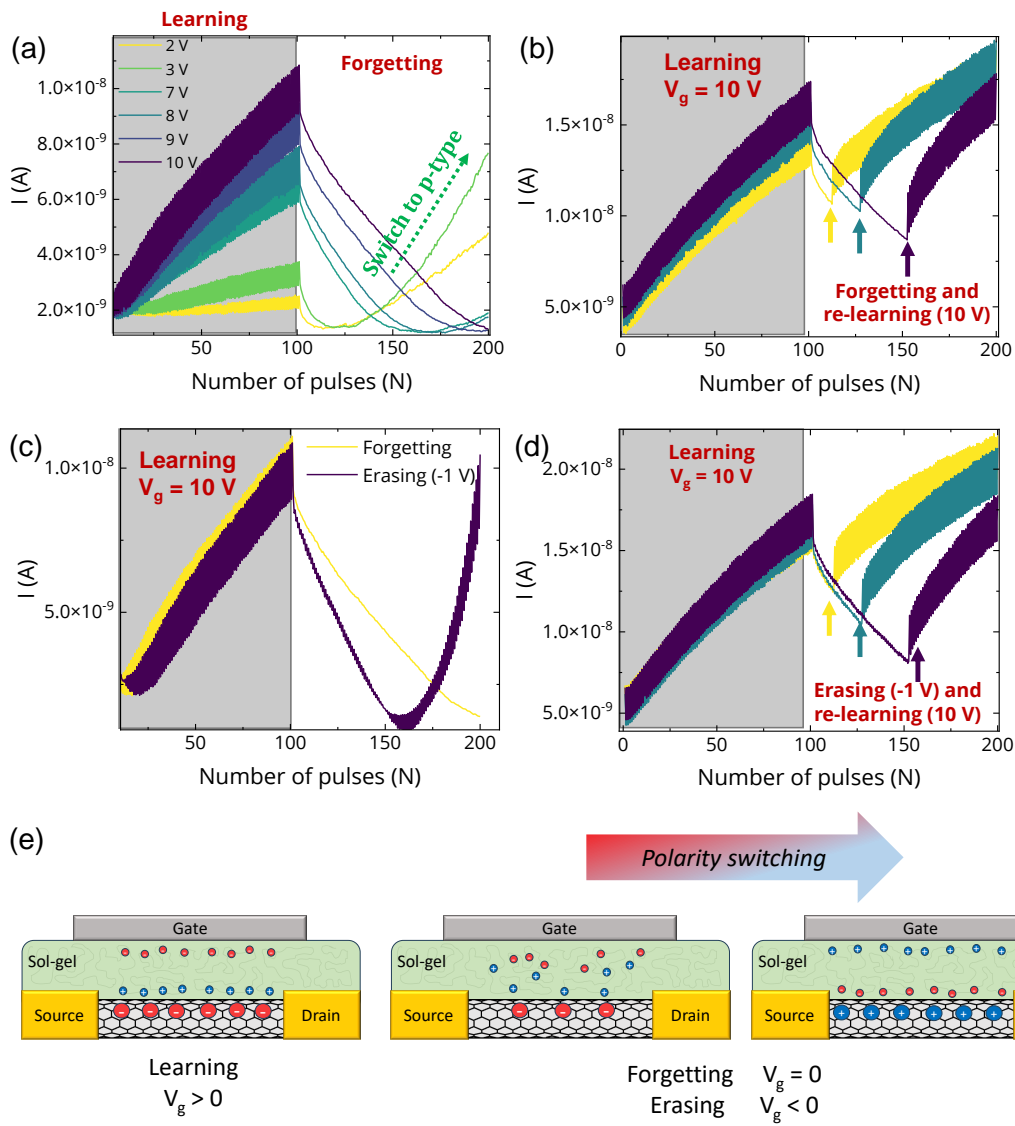

**Figure 6.** Forgetting and erasing in the n-type region of the CNT-FET. (a) Forgetting after different potentiation levels, followed by a switch of polarity. (b) Re-learning after different forgetting delays. (c) Comparison between forgetting and erasing after the same potentiation level, showing a faster polarity switch for the erasing step. (d) Re-learning after different erasing delays. (e) Schematics of the polarity switching effect.

To delve into re-learning following forgetting, we subjected the fabricated samples to a series of 100 consecutive pulses with an amplitude of 10 V, a pulse width of 100 ms, and a fixed time interval between subsequent pulses. This initial phase was aimed at achieving potentiation. Subsequently, we proceeded to potentiate the $I_{ds}$ once again after varying time intervals of forgetting (2 s, 5 s, and 10 s). The findings obtained (Figure 6b), vividly demonstrate that re-learning after forgetting occurs at an accelerated pace compared to the initial potentiation stage. Notably, akin to the p-type region, we observed that longer delays between forgetting and re-learning corresponded to swifter re-learning in the n-type region, albeit not as pronounced as in the p-type region. As the forgotten information was re-learned, the degree of potentiation gradually diminished, ultimately reverted to its normal values.

To scrutinize the relationship between STD and forgetting in the n-type region, we subjected the devices to potentiation during the first 100 positive pulses, and depression during the final 100 pulses, utilizing a pulse amplitude of -1 V, a pulse width of 100 ms, and a time interval of 100 ms. Figure 6c provides a comparative analysis of the STD behavior alongside of forgetting. Our results show that depression occurs faster than forgetting. Moreover, depression has the potential to induce a switch to p-type polarity. Furthermore, the earlier onset in the case of erasing in comparison to forgetting suggests that negative voltages during erasing contribute to cumulative depolarization. However, achieving a complete polarization reset poses a significant challenge due to the tunable threshold voltage and ambipolarity of the device.

To explore the erasing rates associated with different erasing voltages and various potentiation levels, we employed pulse widths of 100 ms and time intervals of 100 ms between consecutive pulses. Since it has proven difficult to achieve exactly the same potentiation level, we compared the erasing voltages by normalizing the $I_{ds}$ values against the maximum potentiation level. The results presented in Figure S9a indicate that higher pulse amplitudes lead to accelerated erasing rates. Furthermore, Figure S9b suggests that higher levels of potentiation do not significantly impact erasing rates. Based on the data presented in Figure S10, it becomes evident that the erasing rate of the neurotransistor is enhanced by large pulse widths, while the time intervals exhibit negligible effects, mirroring the behavior observed in STP in the n-type region. Additionally, as observed in STP in the n-type region, the extremely short pulse widths and time intervals result in an amplified swing of the $I_{ds}$ during erasing, primarily due to capacitive charging effects.

To evaluate the influence of erasing on re-learning, we conducted experiments involving a sequence of -1 V pulses, which would translate to erasing for a specific duration, followed by 100 pulses of 10 V amplitude with a width of 100 ms and a 100 ms time interval. We assessed the impact of 2 s, 5 s, and 10 s delays in re-learning (Figure 6d). Our results indicate that re-learning after erasing exhibits similar behavior to re-learning after forgetting in the n-type region.

**Conclusion**

Neuromorphic computing, a rapidly evolving field that aims at emulating the complex functioning of the human brain, holds tremendous promise for developing highly efficient and scalable computing systems. To further advance this field and overcome the limitations of existing complementary metal-oxide semiconductor solutions, it is crucial to explore novel materials and devices. This study focuses on the fabrication of ambipolar neurotransistors utilizing carbon nanotube field-effect transistors coated with solution-based ion-doped sol-gel silicate film as polarizable gate material that mimics the plasticity observed in synapses. Neurotransistors exhibited ambipolar behavior under slow voltage sweeping speeds, resulting in reduced electron conduction. The study also investigated the learning, forgetting, and erasing behavior of the device, demonstrating short-term depression and short-term potentiation in both the n-type and p-type regions. Our findings reveal that the p-type region exhibited a high ON/OFF ratio, high erasing rate, high learning rate, and easily adjustable parameters. Despite the device's slow switching speeds, its ambipolar behavior and short-term characteristics set it apart from other novel devices. In contrast, the n-type region of the neurotransistor exhibited a low ON/OFF ratio, low learning rate, tunability, and unstable polarization during erasing and forgetting, making it suboptimal for neuromorphic applications. However, the polarity switch can be the base for integrate-and-fire models that bring new functionalities to the device upon further investigation.

Various fabrication techniques were explored to improve device behavior, focusing on varying baking temperatures, air plasma treatments, proportions in precursor ingredients, or the addition of passivation layers. These led to the formation of defective sol-gels or loss of ambipolar characteristics. Consequently, it is necessary to investigate various sol-gel protocols to enhance the adhesion and formation of thinner films with better gate coupling without compromising the memory effects, originating from hysteretic transfer characteristics. Additionally, improvement of ambipolar behavior can be achieved by employing less defective carbon nanotubes. Alternatively, exploration of reconfigurable FETs, such as those utilizing bottom-up silicon nanowires as semiconductors,[34] could be pursued to avoid uncontrolled polarity switching during forgetting and erasing.

The short-term behavior of the device holds promise for adaptable machine learning applications that require real-time data processing. Gas sensing is an example of a suitable application for this neurotransistor. FET-type gas sensors, known for their compact size, ease of integration, and multifunctionality, have gained significant attention.[17] CNTs have also demonstrated ultrasensitivity for gas sensing applications.[17] The combination of carbon nanomaterial-based electronics and machine learning techniques with the requirement to process large amounts of data is an important point of attention in current gas sensing research.[35,36] Ambipolar neuromorphic transistors based on carbon nanotube FETs offer unique advantages over traditional unipolar transistors, making them promising candidates for gas sensing. Their ability to involve two types of charge carriers would enable highly selective sensing, leading to improved discrimination between different gases. They can react double to various gases according to their oxidative or reductive properties, such as $NO_2$, $NH_3$, $H_2S$, and $SO_2$.[37] The excellent compatibility of CNTs with functional materials can further enhance the sensitivity of these sensors, resulting in improved gas detection and signal amplification. Additionally, neuromorphic properties provide simultaneous sensing, storage, and processing capabilities.[38]

**Materials and methods**

*Fabrication of CNT-FETs*

The source and drain electrodes were fabricated following a standard photolithography process and thermal evaporation of Au with a chromium adhesion layer on a silicon wafer with 500 nm oxide layer. First, the samples were put in a sonication bath in acetone for 5 minutes, followed by a sonication bath in isopropanol for 5 minutes, and then rinsed with deionized water. After that, the samples were dried under a stream of nitrogen. Next, a layer of TI Prime was used as an adhesion promoter followed by spin-coating of the AZ5214E Image Reversal photoresist (MicroChemicals GmbH, Germany) using an RC8 coater (Karl Süss, Germany) at 4000 rpm for 40 s to achieve a thickness of 1.5 μm. The samples were baked at 110°C for 60 s. Then, contact lithography UV exposure was performed after chip alignment for 2.5 s. After that, the samples were subjected to photoresist development in AZ726 MIF developer for 60 s and cleaned with deionized water for 1 min. Subsequently, a 5 nm Cr adhesion layer was evaporated onto the substrate, followed by 50 nm of Au using physical vapor deposition. The chip was then submerged in acetone and placed on a shaker for 15-20 min for lift-off. Finally, the samples were rinsed with acetone, isopropanol, and deionized water. The resultant electrode had a channel length of 20 μm and a width of 100 μm.

Deposition of CNTs was done by the alternating current (AC) dielectrophoresis (DEP) process. Sc-SWCNTs were obtained with diameters ranging from 1.2 to 1.7 nm and lengths up to 4 μm from Sigma-Aldrich (Darmstadt, Germany). 0.2 mg of the sc-SWCNTs were dispersed in 12.5 mL of N-methyl pyrrolidone (NMP) by tip sonication for 2 hours at 30% total power in an ice bath, followed by a 1:4 dilution in NMP and re-sonication for 10 minutes under the same conditions. To remove nanotube agglomerations and titanium oxide particles that may have been generated during the sonication process, the resulting dispersion was centrifuged for 30 minutes at 15,000 rpm using a Universal 320 (Hettich, Tuttlingen, Germany). 100 μL of sc-SWCNT solution was deposited onto the electrode region of the substrates and an AC voltage of 10 V peak to peak at a frequency of 30 Hz was applied for 1 minute using a Sony-Tektronix AFG320 function generator to the drain-source electrodes while monitoring signals with a Tektronix TDS3014B oscilloscope. The remaining solution was rinsed with deionized water and then dried using a nitrogen gun.

To minimize the interaction between the drain and source electrodes and the gate electrode, SU8-5 negative epoxy photoresist was used. The SU8-5 negative epoxy photoresist was spin-coated at 3000 rpm for 30 s to achieve a thickness of 5 μm. The samples were then baked at 65°C for 2 minutes and at 95°C for 5 minutes. Then, contact lithography UV exposure was performed for 5 s. The samples were then baked at 65°C for 1 minute and at 95°C for 2 minutes. After that, the samples were subjected to photoresist development in mr600 developer for 2 minutes and cleaned with isopropanol. Finally, the samples were hard-baked at 110°C for an hour.

*Preparation of sol-gel based silicate film and top-gate electrode*

The sol-gel was deposited following a protocol reported in our previous work.[10] The next protocol was followed for 2 μm ion-doped sol-gel silicate film deposition. First, 2.4 mg of nickel(II) chloride hexahydrate ($Cl_2Ni·6H_2O$) and 1.7 mg of copper(II) chloride dehydrate ($Cl_2Cu·2H_2O$) were dissolved in 467 μL of deionized water. Then, 600 μL of trimethoxymethylsilane (MTMS) and 900 μL of tetramethyl orthosilicate (TMOS) precursors were added to the metal salt solution. Next, 33 μL of 0.1 M hydrochloric acid (HCl) was added to lower the pH of the reaction. The glass vial containing the "sol" was vortexed for 2 min and then placed in an ultrasonic bath for 10 min. "Sol" was then passed through a 200 nm pore-size filter to remove large particles. The next protocol was followed for 234 nm ion-doped sol-gel silicate film deposition. First, 4.38 mg of nickel(II) chloride hexahydrate and 3.1 mg of copper(II) chloride dehydrate were dissolved in 100 μL of deionized water. Then, 100 μL of MTMS and 240 μL of TMOS precursors were added to the metal salt solution. Next, 33 μL of 0.1 M HCl was added to lower the pH of the reaction. The glass vial containing the "sol" was vortexed for 2 min and then placed in an ultrasonic bath for 10 min. "Sol" was aged for 6 hours and then passed through a 200 nm pore-size filter to remove large particles.

The next protocol was followed to deposit the sol-gel. CNT-FETs were first cleaned in acetone sonication bath for 2 minutes, then in isopropanol sonication bath for 2 minutes, and then rinsed with deionized water and dried using nitrogen gun. After that, they were annealed at 120°C for 10 minutes to remove any residual solvent or water.[39] Then, 100 W air plasma cleaning was applied for 2 seconds to remove smaller organics and reduce the contact angle, hence enhancing the surface hydrophilicity for sol adhesion. Next, 300 μL of the filtered sol was spin-coated on a 1 cm x 1 cm electrode area of the cleaned chip at 7000 rpm for 60 s. The sol-coated chip was dried in a vacuum oven under 100°C for 24 hours to get a uniform gel formation. Finally, a 70 nm top gate electrode layer of silver was evaporated onto the substrate through a shadow mask using physical vapor deposition.

*Measurement setup*

Measurements were conducted using MATLAB 2016B (Mathworks) software and Keithley 2604B Source Measurement Unit (SMU). To generate the transfer curves, gate voltages were swept at varying speeds, spanning from 0 V to 10 V and subsequently transitioning to -10 V before returning to the initial voltage of 0 V applying a constant $V_{SD}$ of 10 mV. In order to evaluate the effects of erasing (depression), learning (potentiation), and forgetting, different pulse schemes with varying frequencies and amplitudes were employed. Measurements were done at the end of each pulse. A positive 10 V reset voltage was applied again to show the memory effects on the electron conduction polarity of the sol-gel modified CNT-FETs, but the input polarities of the pulses were changed.

**Acknowledgments**

We acknowledge the financial support by the Federal Ministry of Education and Research of Germany in the program of "Souverän. Digital. Vernetzt." joint project 6G-life, project number: 16KISK001K. We thank Dr. Eunhye Baek for efficient technical troubleshooting.

Supplementary information to:

Carbon nanotube neurotransistors with ambipolar memory and learning functions

Ertürk Enver Yildirim[1], Luis-Antonio Panes-Ruiz[1] ORCID ID: 0000-0002-3007-8840, Pratyaksh Yemulwar[1], Ebru Cihan[1] ORCID ID: 0000-0002-1747-3838, Bergoi Ibarlucea[1*] ORCID ID: 0000-0002-9899-1409, Gianaurelio Cuniberti[1*] ORCID ID: 0000-0002-6574-7848

[1]Institute for Materials Science and Max Bergmann Center for Biomaterials, Dresden University of Technology, Dresden (Germany)

*Bergoi Ibarlucea: bergoi.ibarlucea@tu-dresden.de; Gianaurelio Cuniberti: gianaurelio.cuniberti@tu-dresden.de

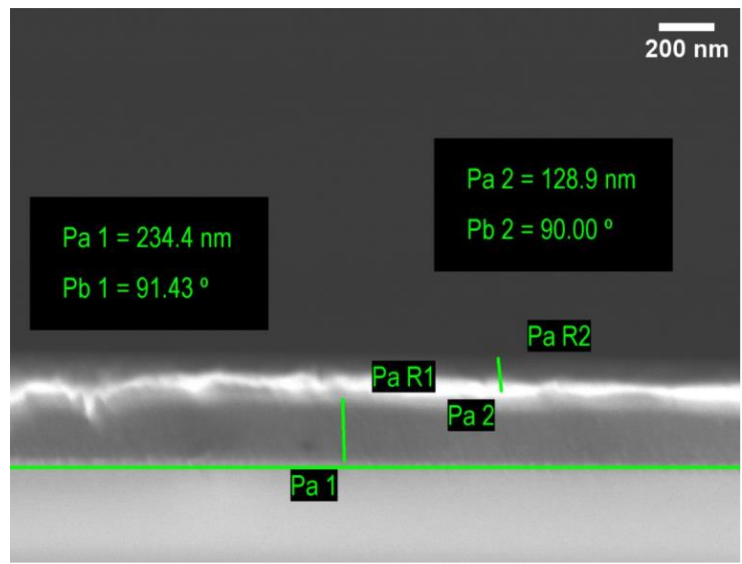

**Figure S1.** Scanning electron microscopy (SEM) of a thinner sol-gel obtained by dilution of the precursor mixture. Pa1 = sol-gel layer. Pa2 = Ag layer.

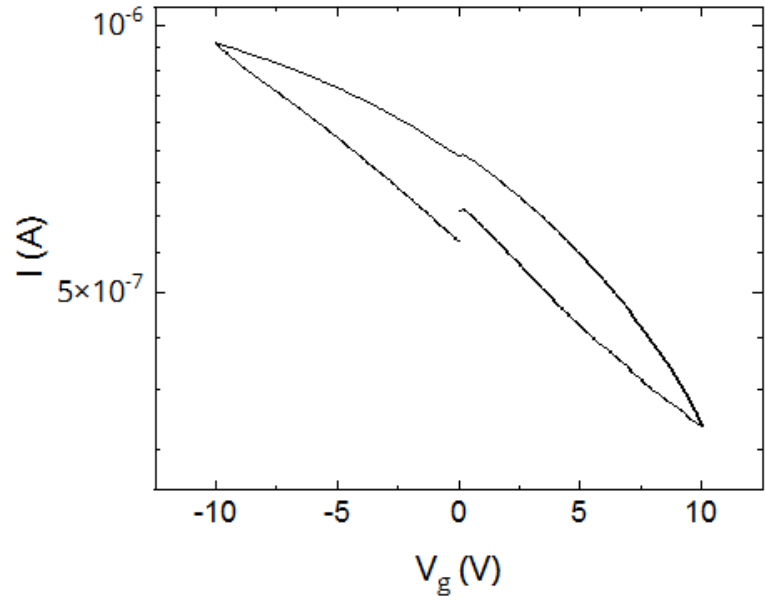

**Figure S2.** Transfer characteristics of the thin sol-gel obtained by dilution of the precursor mixture.

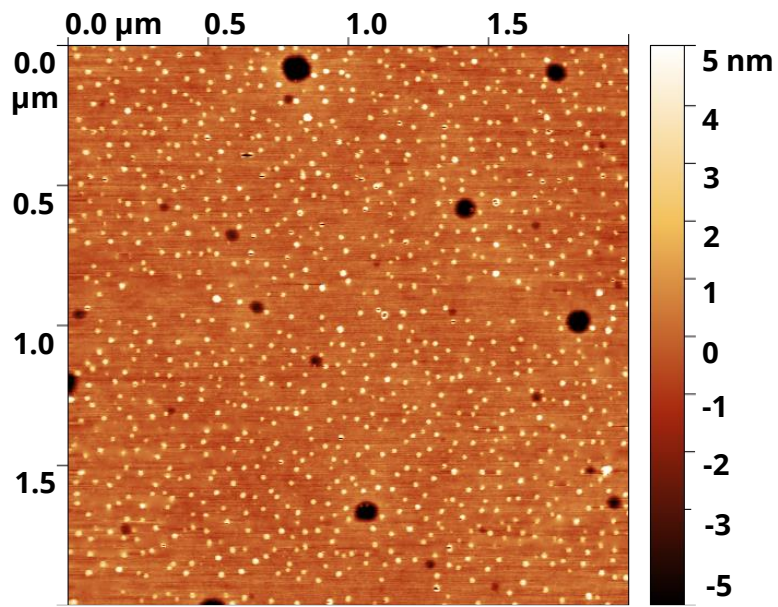

**Figure S3.** Atomic Force Microscopy (AFM) image of the thin sol-gel obtained by dilution of the precursor mixture.

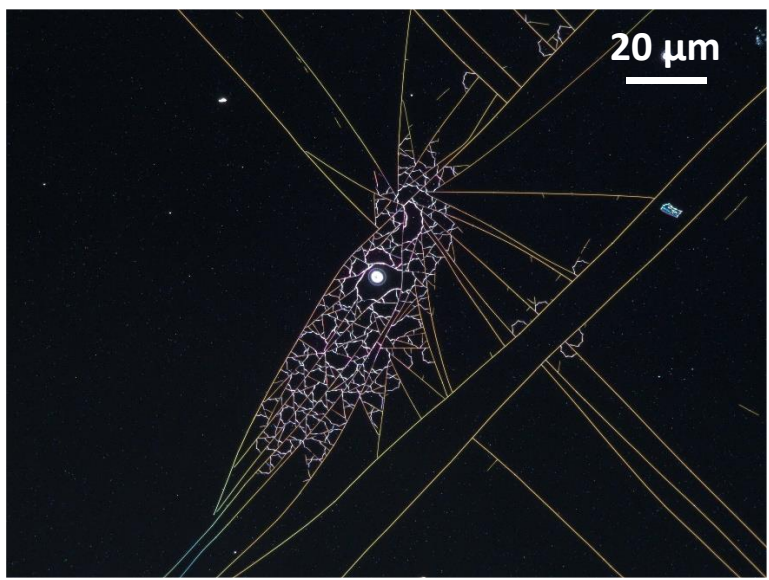

**Figure S4.** Optical microscopy of cracked sol-gel layer fabricated under 200 °C heating conditions.

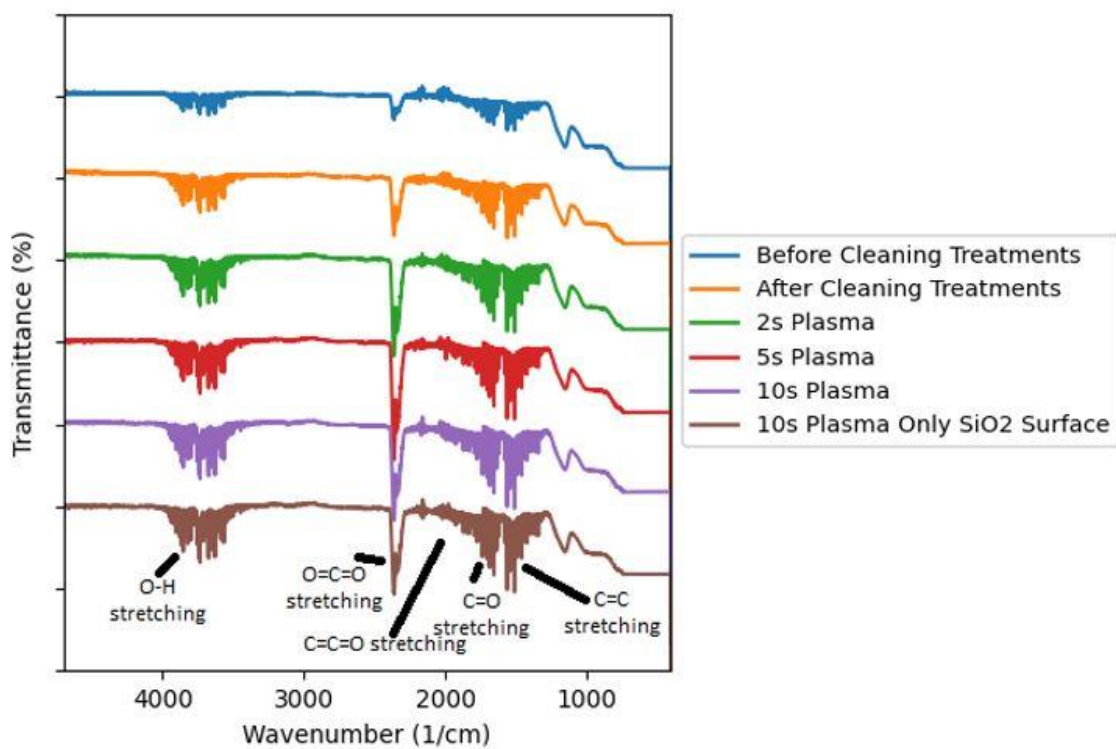

**Figure S5.** Fourier Transform Infrared Spectroscopy of carbón nanotube samples treated with air plasma durations of up to 10 seconds before sol-gel deposition.

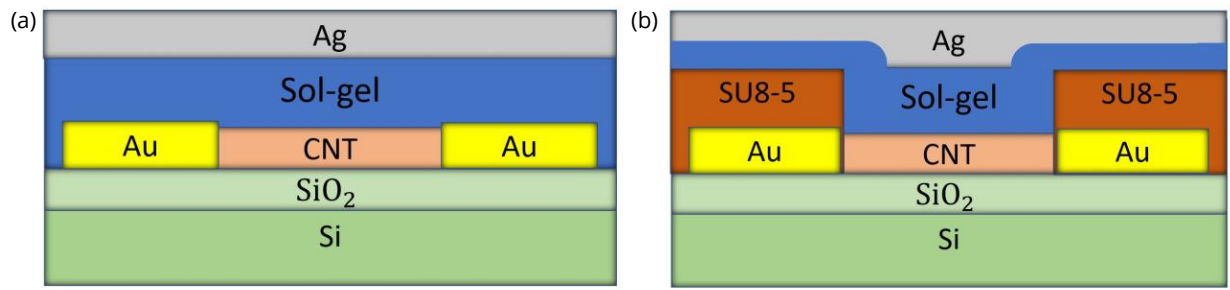

**Figure S6.** Sketch of the cross-section of the fabricated neurotransistor (a) without SU8-5 and (b) with SU8-5 passivation.

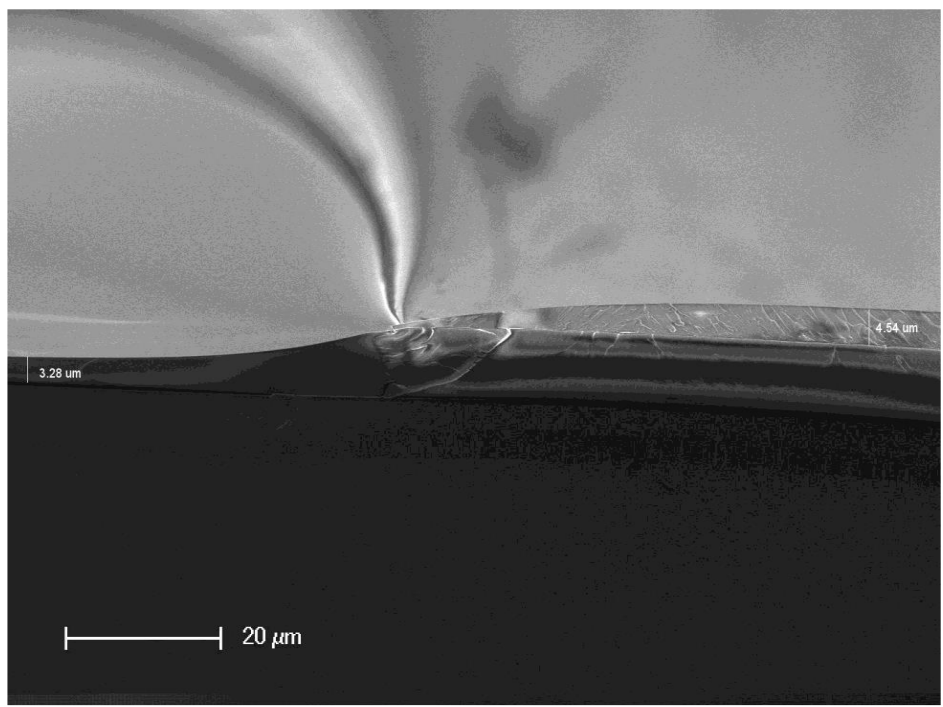

**Figure S7.** SEM image of a sol-gel film deposition at the edge of SU8-5 passivation.

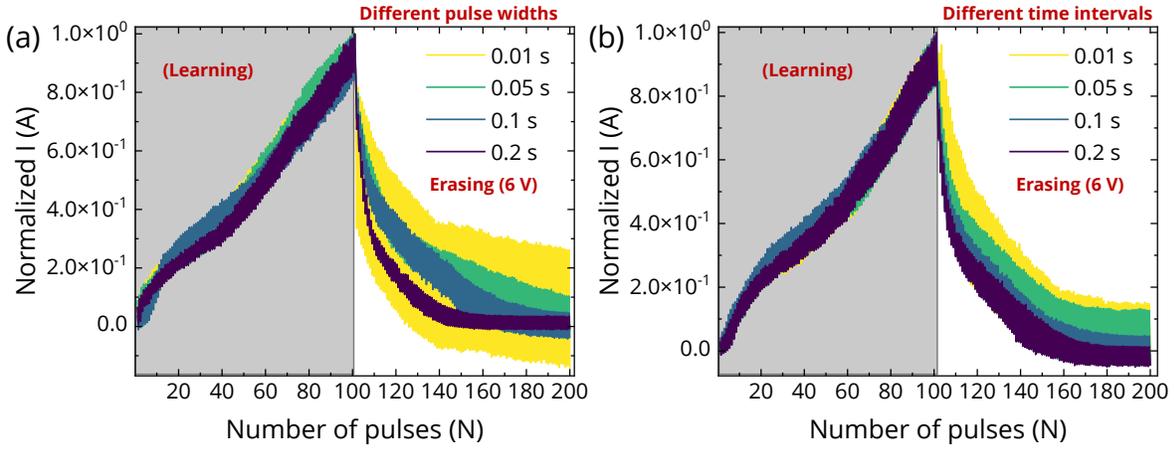

**Figure S8.** Normalized current resulting from learning and erasing in the p-type region (a) with different pulse widths, and (b) with different time intervals between consequent pulses.

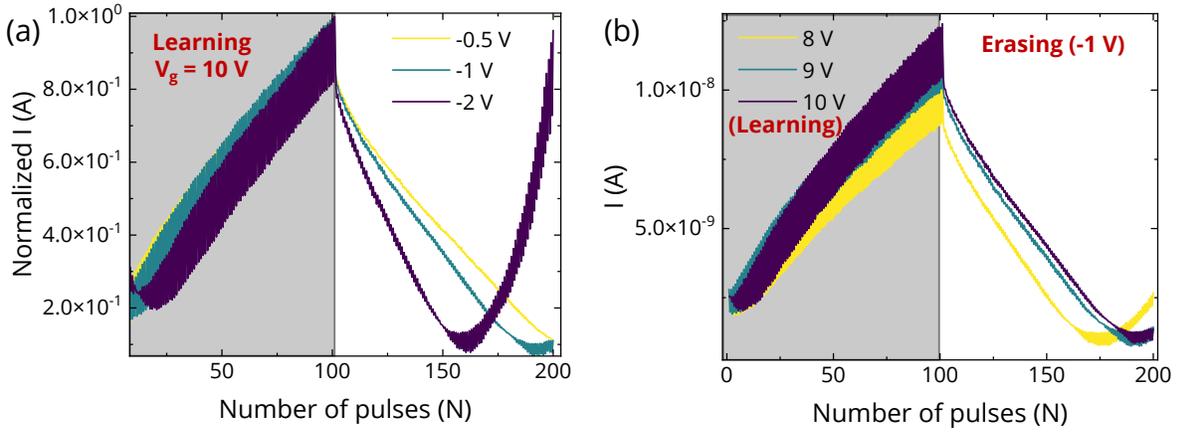

**Figure S9**: Comparison of erasing after potentiation in the n-type región: (a) from different potentiation levels, and (b) erasing with different voltages (normalized current).

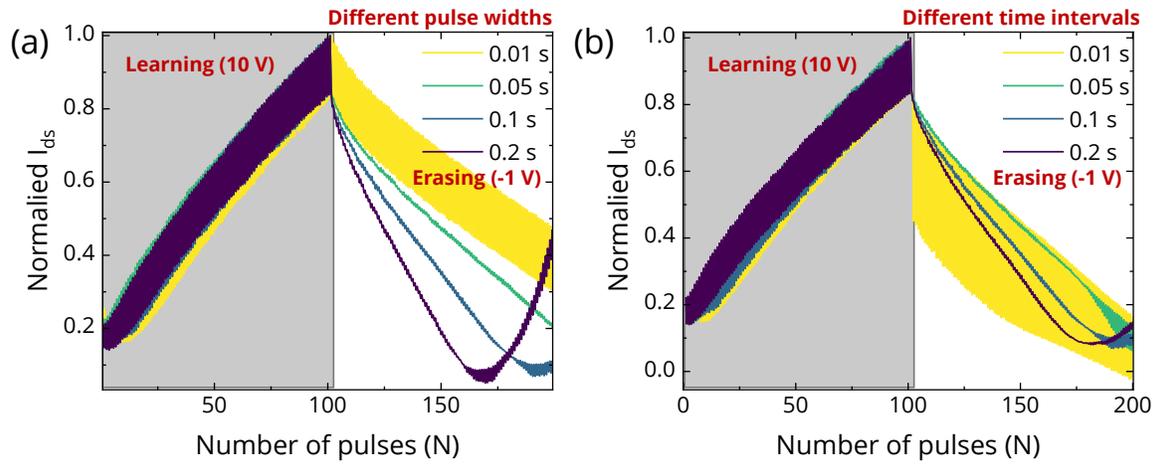

**Figure S10**: normalized $I_{ds}$ short-term depression (STD) (a) with different pulse widths, and (b) with different time intervals between consequent pulses in the n-type región.